# Stability of Degree Heterogeneous Ecological Networks

Gang Yan, Neo D. Martinez, and Yang-Yu Liu

**A classic measure of ecological stability describes the tendency of a community to return to equilibrium after small perturbation. While many advances show how the network structure of these communities severely constrains such tendencies, few if any of these advances address one of the most fundamental properties of network structure: heterogeneity among nodes with different numbers of links. Here we systematically explore this property of "degree heterogeneity" and find that its effects on stability systematically vary with different types of interspecific interactions. Degree heterogeneity is always destabilizing in ecological networks with both competitive and mutualistic interactions while its effects on networks of predator-prey interactions such as food webs depend on prey contiguity, i.e., the extent to which the species consume an unbroken sequence of prey in community niche space. Increasing degree heterogeneity stabilizes food webs except those with the most contiguity. These findings help explain previously unexplained observations that food webs are highly but not completely contiguous and, more broadly, deepens our understanding of the stability of complex ecological networks with important implications for other types of dynamical systems.**

Understanding the intricate relationship between the structure and dynamics of complex ecological systems has been one of the key issues in ecology [1-4]. Equilibrium stability of ecological systems, a measure that considers an ecological system stable if it returns to its equilibrium after a small perturbation, has been a central research topic for over four decades [1, 5-15]. Empirical observations suggest that communities with more species are more stable, i.e., a positive diversity-stability relationship [16]. Yet, these intuitive ideas were challenged by the pioneer work of May [1, 2], who rigorously proved a negative diversity-stability relationship using linear stability analysis on randomly constructed ecological communities. The contradiction between these findings forms the eminent "diversity-stability debate" [6, 7].

May's seminal work [1, 2] considered community matrices $M$ of size $S \times S$, where $S$ is the number of species, the off-diagonal elements $M_{ij} \equiv \left.\frac{\partial f_i(x)}{\partial x_j}\right|_{x=x^*}$ captures the impact that species $j$ has on species $i$ around a feasible equilibrium point $x^*$ of an unspecified dynamical system $\dot{x}(t) = f(x(t))$ describing the time-dependent abundance $x(t)$ of the $S$ species. Since empirical parameterization of the exact functional form of $f(x(t))$ is extremely difficult for complex ecological systems, May considered $M_{ij}$'s are randomly drawn from a distribution with mean zero and variance $\sigma^2$ with probability $C$ and are 0 otherwise. Hence $\sigma$ represents the characteristic interaction strength and $C$ is the ratio between actual and potential interactions in the ecological system. For simplicity, the diagonal elements are chosen to be the same, $-1$, representing the intrinsic dampening time scale of each species so that if disturbed from equilibrium it would return with such a dampening time by itself. May found that for random interactions drawn from a normal distribution $\mathcal{N}(0, \sigma^2)$, a randomly assembled system is stable (i.e., all the eigenvalues of the community matrix $M$ have negative real parts) if the so-called 'complexity' measure $\sigma\sqrt{CS} < 1$. This implies that more complexity (e.g., due to larger $C$ or $S$) tend to destabilize community dynamics [1, 2].

May's result continues to be influential almost four decades later is not because that complex ecological systems have to be unstable, but because real ecological system must have some specific structures that allow them to be stable despite their complexity [4]. In other words, nature must adopt what May called "devious strategies" to cope with this diversity-stability paradox. One of such strategy is the existence of particular interspecific relationships observed in nature, e.g., predator-prey, competition, and mutualism. Recently Allesina and Tang refined May's result and provided analytic stability criteria for all these interspecific interaction types [14]. They found remarkable differences between predator-prey interactions, which are stabilizing, and mutualistic and competitive interactions, which are destabilizing. These newer findings allow many different strategies employed by nature to be more powerfully tested with the refined stability criteria as a reference point.

**Degree heterogeneity**

The above results rely on a key assumption that the underlying network structure is almost completely random. Indeed, the construction of the community matrix $M$ follows almost the same procedure as the classical Erdös-Rényi random graph model [17]. However, just like many other real-world complex systems, the underlying networks of ecological systems are far from random. Instead, they often display non-trivial topological features, e.g., degree heterogeneity (the variation among less and more connected nodes) [18-21], nestedness (the tendency for the links of specialists to be subsets of the links of generalists) [22], and modularity (the degree of compartmentalization of the networks) [23]. It has been recently shown that the network architecture favoring stability fundamentally differs between trophic and mutualistic networks [12]. For example, a highly connected and nested architecture promotes community stability in mutualistic networks [11], whereas the stability of trophic networks is enhanced in modular and weakly connected architectures.

Despite these and other remarkable results, much about the relationship between network structure and stability of complex ecological systems remains unknown. One fundamental issue concerns the impact of degree heterogeneity. Typically the degree heterogeneity of a network can be represented by $\xi \equiv \langle k^2 \rangle / \langle k \rangle^2$, where $\langle k \rangle = 1/S \sum_{i=1}^{S} k_i$ is the mean degree of the network, $k_i$ is the degree of species $i$ (i.e., the total number of incoming and outgoing connections species $i$ has), and $\langle k^2 \rangle = 1/S \sum_{i=1}^{S} k_i^2$ is the second moment of the species degree distribution $P(k)$. Note that $\xi > 1$. Also, the higher the degree heterogeneity, the larger the value of $\xi$. A key advance in understanding complex networks over the last decade has been how powerfully degree heterogeneity affects many network properties and dynamical processes, from error and attack tolerance [24, 25], to graph spectra [26], epidemic spreading [27], interdependent fragility [28], and controllability [29]. It is reasonable to expect that degree heterogeneity would affect stability of complex ecological systems as well. To approach this issue in a systematic fashion, we follow May's model-independent framework [1], helping us avoid the difficulty of parameterizing the exact dynamics of complex ecological systems.

**Degree heterogeneous network with random interactions**

For an ecological system with random interactions, we generate its underlying network topology first, using three different network models (see SI Sec. I for details): multi-modal [30], Erdös-Rényi (ER) [17], and scale-free (SF) [31], with given mean degree $\langle k \rangle$ and effective connection probability $C = \langle k \rangle/(2(S-1))$. Then we construct the community matrix $M$ as follows: (1) Set all the diagonal elements $M_{ii} = -d$; (2) Draw the off-diagonal element $M_{ij}$ from a normal distribution $\mathcal{N}(0, \sigma^2)$ whenever there is a link from species $j$ to species $i$. The real part of $M$'s most positive eigenvalue is given by $\mathrm{Re}(\lambda_\mathrm{m}) = \sigma\sqrt{\xi \langle k \rangle/2} - d = \sqrt{\xi(S-1)C} - d$ (see SI Sec. III for derivation). $\mathrm{Re}(\lambda_\mathrm{m})$ has to be negative to ensure the equilibrium stability, yielding the stability criterion

$$\sqrt{\xi(S-1)C} < d. \qquad (1)$$

Apparently, any factor that increases (or decreases) $\mathrm{Re}(\lambda_\mathrm{m})$ will destabilize (or stabilize) the ecological system, respectively. Increasing $\xi$ will certainly destabilize the ecological system. Note that for random networks generated by the Erdös-Rényi model, $\xi \to 1$ for large $\langle k \rangle$. Hence Eq.(1) naturally recovers May's classical result for large $S$ [1].

Fig. 1a shows the impact of degree heterogeneity on the stability of ecological systems with random interactions. The underlying architectures are constructed from different network models with tunable degree heterogeneity $\xi$ (SI Sec. I). We find that when the complexity $\sigma\sqrt{CS}$ is fixed, $\mathrm{Re}(\lambda_\mathrm{m})$ grows monotonically as $\xi$ increases, implying that larger $\xi$ always destabilizes an ecological system with random interactions. Moreover, our numerical results suggest that $(\text{Re}(\lambda_m) + d) \propto \xi^{1/2}$, which agrees well with our analytical prediction (SI Sec. III). This finding can be further illustrated by the distribution of $M$'s eigenvalues. We show that for increasing $\xi$, the radius of the cycle encompassing all the eigenvalues becomes larger, hence $\text{Re}(\lambda_m)$ increases, destabilizing the ecological system. Interestingly, we also find that as $\xi$ increases the eigenvalue distribution becomes more non-uniform with very high density of eigenvalues concentrated around the center of the circle (Figs. 1b-d).

**Degree heterogeneous network with predator-prey interactions**

For ecological systems with predator-prey interactions, i.e., whenever $M_{ij} > 0$ then $M_{ij} < 0$, we generate the underlying network using five different models (see SI Sec. I for details): multimodal, Erdös-Rényi, scale-free, cascade [32], and niche model [33], with given mean degree $\langle k \rangle$. When there is a directed edge from species $i$ to $j$, we draw the off-diagonal element $M_{ij}$ from the half-normal distribution $|\mathcal{N}(0, \sigma^2)|$ and $M_{ji}$ from $-|\mathcal{N}(0, \sigma^2)|$. We still set all the diagonal elements $M_{ii} = -d$. As shown in Figs. 2a and 2b, for simple model networks (3-modal, ER and SF), large degree heterogeneity hampers the stability of predator-prey ecological systems, yet moderate heterogeneity is stabilizing. This non-monotonic behavior can be further illustrated by the eigenvalue distribution (Fig. 2c). We find that as degree heterogeneity $\xi$ increases, not only does the eigenvalue distribution become non-uniform but also, the boundary that encompasses all eigenvalues changes from ellipse to bow-tie shape (Fig. 2c). This alteration of boundary shape induces the non-monotonic behavior of $\text{Re}(\lambda_m)$ for varying $\xi$.

In empirical food webs there exist trophic hierarchy and niche structure [34]. To reproduce these structural properties we employ the widely used cascade [32] and niche [33] models. When the number of species $S$ and the connection probability $C$ are fixed, we tune the degree heterogeneity $\xi$ to assess its impact on stability. As shown in Fig. 3a, for networks generated by the niche model degree heterogeneity $\xi$ destabilizes predator-prey ecological systems, in contrast to networks generated by the cascade model for which degree heterogeneity is stabilizing (Fig. 3b). We also calculate the eigenvalue distributions of the community matrix associated with the niche model, finding that there is no drastic shape alteration of the boundary encompassing all the eigenvalues (Fig. 3c).

A primary distinction between the cascade and niche models is *prey contiguity,* which describes the tendency for species to consume a contiguous sequence of prey in tropic niche space of the whole community [33, 34] (Fig. 4a). The remarkable difference between Fig. 3a and Fig. 3b prompts us to systematically explore the effect of prey contiguity. We adopt the relaxed niche model [35] to generate underlying networks with tunable prey contiguity $g$ such that $g = 0$ (or $g = 1$) corresponds to the original cascade (or niche) model, respectively. We find that, indeed, the impact of degree heterogeneity on stability depends on the prey contiguity $g$ (Fig. 4b). In particular, degree heterogeneity is stabilizing for $g < g^*$ while destabilizing for $g > g^*$, where $g^* \approx 0.85$ (Fig. 4c). Interestingly, the contiguity of most empirical food webs [35] lies in the regime where degree heterogeneity favors community stability, suggesting that the existing degree heterogeneity in real-world food webs might have been shaped by stability. To test this hypothesis, for each empirical food web we generate a model network whose degree heterogeneity and prey contiguity are equal to that of the empirical network. As shown in Fig. 4d the stability of empirical food webs are well approximated by the corresponding model networks, indicating that degree heterogeneity and prey contiguity together may largely determine the stability of predator-prey networks (See also SI Sec IV). To our knowledge, there had previously been no explanation for the prominent empirical finding that feeding niches of predators in real food webs are close to contiguous but not completely so (i.e., the prey contiguity $g$ is close to but not exactly 1). Our result that in the presence of degree heterogeneity complete contiguity ($g = 1$) is destabilizing food webs may go a long way in explaining this curious aspect of ecological networks.

The finding that a slight reduction of prey contiguity leads to such a qualitatively different stabilizing effect deserves further studies. Our preliminary results suggest that the network becomes more stable if, for each species, we rewire a fraction of its links to connect the species below its feeding niche. Otherwise, the network becomes more unstable and the degree heterogeneity is destabilizing. In other words, links that connect species more closely to the energy source (plants) appear to be driving the stabilizing effect (see SI Sec. IV for details).

**Heterogeneous network with mixed interactions of competition and mutualism**
For ecological systems with a mixture of competitive and mutualistic interactions where $M_{ij}$ and $M_{ji}$ always have the same sign, we construct the community matrix $M$ following a similar approach as the case of predator-prey interactions. We find that for all multi-modal (Fig. 5a), Erdös-Rényi and scale-free (Fig. 5b) model networks the $\text{Re}(\lambda_m)$ grows monotonically as we increase $\xi$, implying that degree heterogeneity is always destabilizing for competitive and mutualistic systems. This is further demonstrated by the eigenvalue distribution of the community matrix M (Fig. 5c). Similar to the predator-prey ecological systems, we find that when the degree heterogeneity $\xi$ increases, not only the eigenvalue distribution becomes non-uniform but also, the boundary that encompasses all the eigenvalues changes from ellipse to bow-tie shape (Fig. 5c). Despite the alteration of the eigenvalue distribution, the real parts of the eigenvalues always expand as $\xi$ increases. This explains the monotonic impact of degree heterogeneity on the stability of ecological systems with a mixture of competitive and mutualistic interactions. Note that for symmetric networks with only mutualistic interactions the degree heterogeneity is also destabilizing [36].

**Discussion**

In summary, with extensive numerical and analytical calculations, we find that for ecological networks with random interspecific interactions or a mixture of competitive and mutualistic interactions, the degree heterogeneity always destabilizes ecological systems. For ecological networks with predator-prey interactions (e.g., food webs), the impact of degree heterogeneity on stability depends on the prey contiguity. When the prey contiguity is not very high, which is true for most empirical food webs, degree heterogeneity tends to stabilize predator-prey systems.

The structure of ecological networks has been recognized as one key ingredient contributing to the coexistence between high complexity and stability in real ecological systems [4, 6, 7, 37]. Our results demonstrate that, depending on the type of interactions, degree heterogeneity of ecological networks has fundamentally different impacts on the community stability. This implies that strong variations in the stability of architectural patterns constrain ecological networks toward different architectures, consistent with previous results [12, 14]. The presented results offer a novel angle to understand how nature produces diverse, yet stable ecological systems. Moreover, since we use the model independent framework and linear stability analysis, our findings are not limited to ecological networks, but instead hold for any system of differential equations resting at an equilibrium point. For example, our results could be relevant in financial [38], organizational [39], and biological [40] systems where competition, cooperation and consumer-resource interactions occur.

1. May, R. M. Will a large complex system be stable? Nature 238, 413–414 (1972).

2. May, R. M. Stability and Complexity in Model Ecosystems (Princeton Univ. Press, 2001).

3. Pascual, M. & Dunne, J. Ecological Networks: Linking Structure to Dynamics in Food Webs (Oxford University Press, 2006).

4. Bascompte, J. Structure and dynamics of ecological networks. Science 329, 765–766 (2010).

5. Pimm, S. L. The complexity and stability of ecosystems. Nature 307, 321–326 (1984).

6. McCann, K. The diversity-stability debate. Nature 405, 228–233 (2000).

7. Ives, A. R. & Carpenter, S. R. Stability and diversity of ecosystems. Science 317, 58–62 (2007).

8. Gross, T., Ebenhh, W. & Feudel, U. Enrichment and foodchain stability: the impact of different forms of predator-prey interaction. J. Theor. Biol. 227, 349–358 (2004).

9. Gross, T., Rudolf, L., Levin, S. A. & Dieckmann, U. Generalized models reveal stabilizing factors in food webs. Science 325, 747–750 (2009).

10. Okuyama, T. & Holland, J. N. Network structural properties mediate the stability of mutualistic communities. Ecol. Lett. 11, 208–216 (2008).

11. Bastolla, U. *et al.* The architecture of mutualistic networks minimizes competition and increases biodiversity. Nature 458, 1018–1020 (2009).

12. Thébault, E. & Fontaine, C. Stability of ecological communities and the architecture of mutualistic and trophic networks. Science 329, 853–856 (2010).

13. Mougi, A. & Kondoh, M. Diversity of interaction types and ecological community stability. Science 337, 349–351 (2012).

14. Allesina, S. & Tang, S. Stability criteria for complex ecosystems. Nature 483, 205–208 (2012).

15. Suweis, S., Simini, F., Banavar, J. R. & Maritan, A. Emergence of structural and dynamical properties of ecological mutualistic networks. Nature 500, 449–452 (2013).

16. Elton, C. S. Ecology of Invasions by Animals and Plants (Univ. of Chicago Press, 2000).

17. Erdös, P. & Rényi, A. On the evolution of random graphs. Publ. Math. Inst. Hung. Acad. Sci. 5, 17–60 (1960).

18. Dunne, J. A., Williams, R. J. & Martinez, N. D. Food-web structure and network theory: The role of connectance and size. Proc. Natl. Acad. Sci. USA 99, 12917–12922 (2002).

19. Jordano, P., Bascompte, J. & Olesen, J. M. Invariant properties in coevolutionary networks of plant-animal interactions. Ecol. Lett. 6, 69–81 (2003).

20. Vázquez, D. P. Degree distribution in plant-animal mutualistic networks: forbidden links or random interactions? Oikos 108, 421–426 (2005).

21. Montoya, J. M., Pimm, S. L. & Solé, R. V. Ecological networks and their fragility. Nature 442, 259–264 (2006).

22. Bascompte, J., Jordano, P., Melian, C. J. & Olesen, J. M. The nested assembly of plant-animal mutualistic networks. Proc. Natl. Acad. Sci. USA 100, 9383–9387 (2003).

23. Krause, A. E., Frank, K. A., Mason, D. M., Ulanowicz, R. E. & Taylor, W. W. Compartments revealed in food-web structure. Nature 426, 282–285 (2003).

24. Albert, R., Jeong, H. & Barabási, A.-L. Error and attack tolerance of complex networks. Nature 406, 378–382 (2000).

25. Cohen, R., Erez, K., ben Avraham, D. & Havlin, S. Resilience of the internet to random breakdowns. Phys. Rev. Lett. 85, 4626–4628 (2000).

26. Dorogovtsev, S. N., Goltsev, A. V., Mendes, J. F. F. & Samukhin, A. N. Spectra of complex networks. Phys. Rev. E 68, 046109 (2003).

27. Pastor-Satorras, R. & Vespignani, A. Epidemic spreading in scale-free networks. Phys. Rev. Lett. 86, 3200–3203 (2001).

28. Buldyrev, S. V., Parshani, R., Paul, G., Stanley, H. E. & Havlin, S. Catastrophic cascade of failures in interdependent networks. Nature 464, 1025–1028 (2010).

29. Liu, Y.-Y., Slotine, J.-J. & Barabási, A.-L. Controllability of complex networks. Nature 473, 167–173 (2011).

30. Valente, A. X. C. N., Sarkar, A. & Stone, H. A. Two-peak and three-peak optimal complex networks. Phys. Rev. Lett. 92, 118702 (2004).

31. Barabási, A.-L. & Albert, R. Emergence of scaling in random networks. Science 286, 509–512 (1999).

32. Cohen, J. E. & Newman, C. M. A stochastic theory of community food webs: I. models and aggregated data. Proc. R. Soc. B 224, 421–448 (1985).

33. Williams, R. J. & Martinez, N. D. Simple rules yield complex food webs. Nature 404, 180–183 (2000).

34. Cohen, J. E. Food Webs and Niche Space (Princeton University Press, 1978).

35. Williams, R. J. & Martinez, N. D. Success and its limits among structural models of complex food webs. Journal of Animal Ecology 77, 512–519 (2008).

36. Feng, W. & Takemoto, K. Heterogeneity in ecological mutualistic networks dominantly determines community stability. Sci. Rep. 4, 5912 (2014).

37. Rohr, R. P., Saavedra, S. & Bascompte, J. On the structural stability of mutualistic systems. Science 345, 6195 (2014).

38. Haldane, A. G. & May, R. M. Systemic risk in banking ecosystems. Nature 469, 351–355 (2011).

39. Saavedra, S., Reed-Tsochas, F. & Uzzi, B. A simple model of bipartite cooperation for ecological and organizational networks. Nature 457, 463–466 (2008).

40. Foster, K. & Bell, T. Competition, not cooperation, dominates interactions among culturable microbial species. Current Biology 22, 1845–1850 (2012).

41. Stouffer, D. B., Camacho, J., Guimerà, R., Ng, C. A. & Amaral, L. A. N. Quantitative patterns in the structure of model and empirical food webs. Ecology 86, 1301–1311 (2005).

**Author Contributions** Y.-Y.L conceived the project. G.Y., N.D.M. and Y.-Y.L. designed and performed the research, analyzed the results, and wrote the manuscript.

**Acknowledgments** We thank Hai-jun Zhou for valuable discussions. Y.-Y.L. gratefully acknowledges the support from the John Templeton Foundation (award #51977).

**Competing Interests** The authors declare that they have no competing financial interests.

**Correspondence** Correspondence and requests for materials should be addressed to Yang-Yu Liu (email: yyl@channing.harvard.edu).

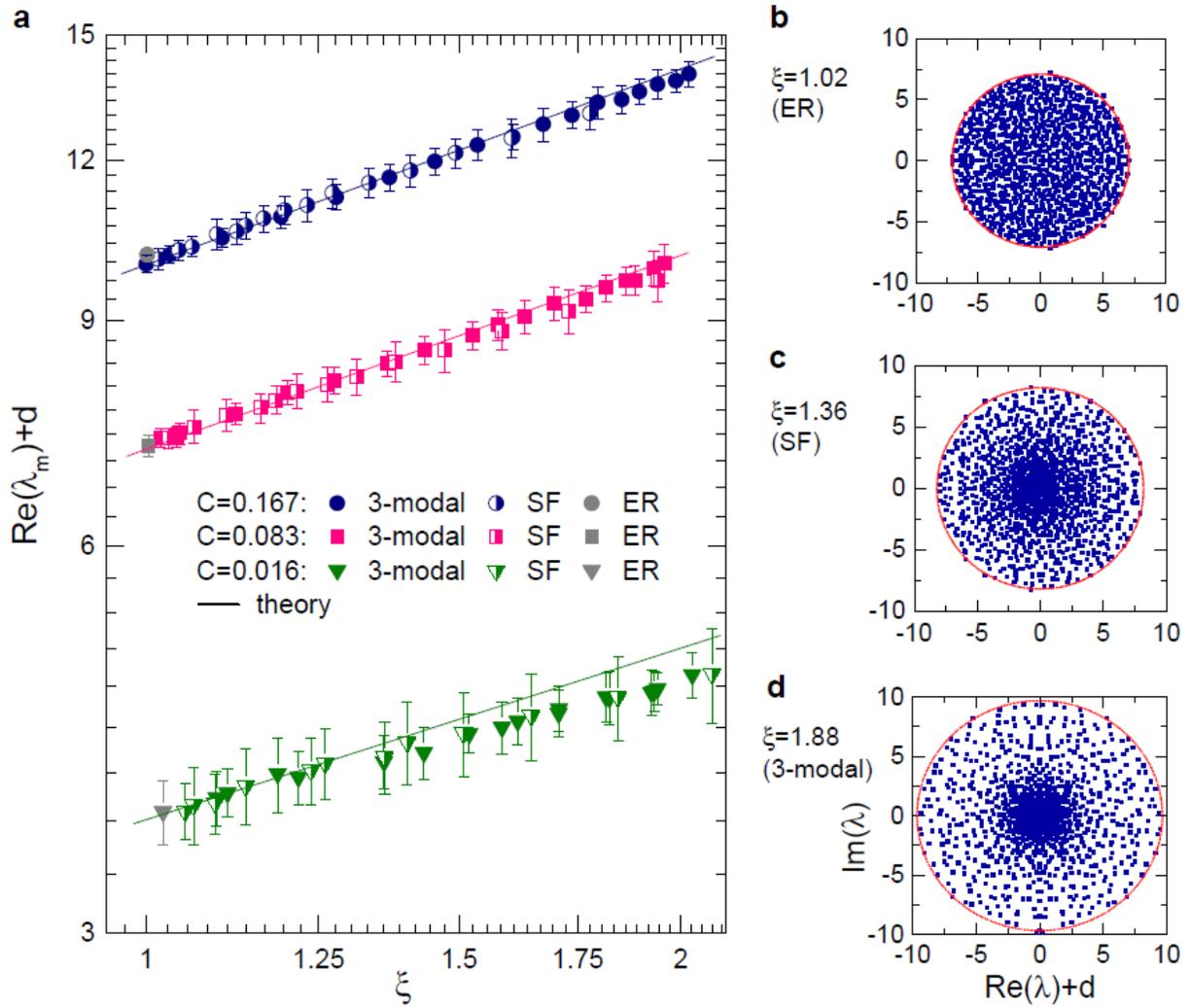

**Figure 1: Stability vs degree heterogeneity for ecological networks with random interactions.** (a) $\text{Re}(\lambda_\text{m}) + d$ as a function of degree heterogeneity, $\xi \equiv \langle k^2 \rangle / \langle k \rangle^2$, in log-log plot. The dots represent the results from numerical simulations on 3-modal, Erdös-Rényi (ER), and scale-free (SF) networks. For 3-modal networks the degree heterogeneity $\xi$ is tuned by the variance of degree sequence (see SI Sec. IA). For scale-free networks $\xi$ is tuned by the power-law exponent of degree distribution (see SI Sec. IC). The network size $S = 1200$, the connection probability $C = 0.167, 0.083, 0.016$ (i.e., the mean degree $\langle k \rangle = 100, 50, 10$), and the strength of edges are drawn from the normal distributions $\mathcal{N}(0, \sigma^2)$ with $\sigma = 1.0$. Each error bar represents the standard deviation of 100 independent runs. The solid lines (all with the slope 1/2) are derived from our analytical results. (**b-d**) The typical distributions of the eigenvalues for Erdös-Rényi ($\xi = 1.02$), scale-free ($\xi = 1.36$) and 3-modal ($\xi = 0.88$) networks. The radius of the boundary circles are $\sigma\sqrt{\xi \langle k \rangle / 2}$.

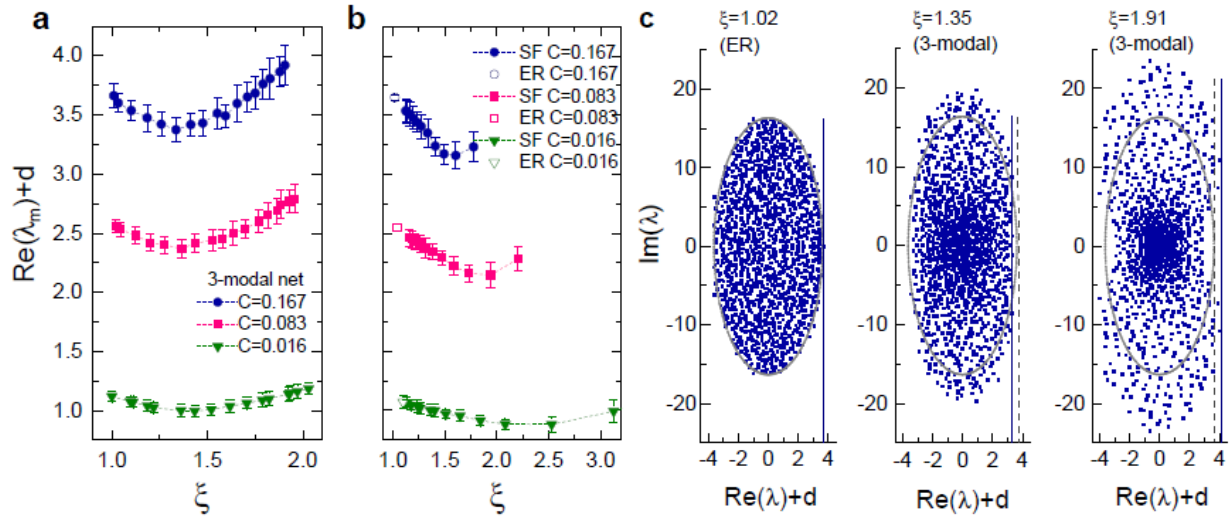

**Figure 2: Stability vs degree heterogeneity for simple model networks with predator-prey interactions. (a,b)** $\text{Re}(\lambda_m) + d$ as a function of the degree heterogeneity $\xi$ for 3-modal, Erdös-Rényi (ER), and scale-free (SF) model networks. $S = 1200$, and $\sigma = 1.0$. **(c)** The eigenvalue distributions of ER ($\xi = 1.02$) and 3-modal ($\xi = 1.35, 1.91$) networks. The ellipse (drawn for reference) is the result for homogeneous networks [14]. Because of the bow-tie shape of the eigenvalue distribution, when $\xi = 0.35$ the real part of the most positive eigenvalue (solid line) is smaller than that of the ellipse (dashed line); the opposite is true when $\xi = 1.91$.

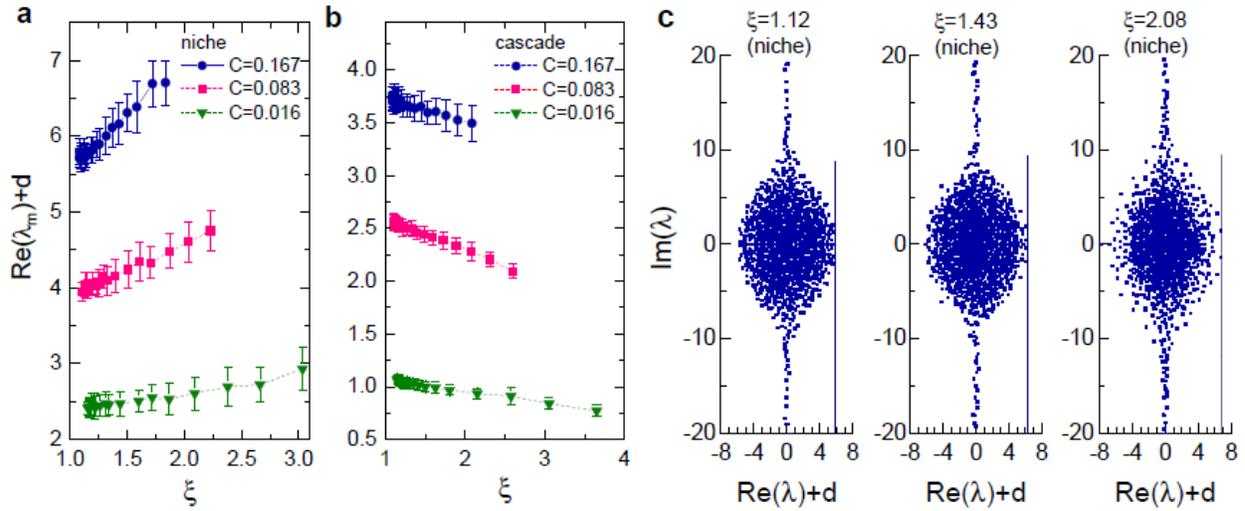

**Figure 3: Stability vs degree heterogeneity for realistic model food webs.** (a,b) $\text{Re}(\lambda_m) + d$ as a function of the degree heterogeneity $\xi$. The networks are generated using the niche [33] and cascade [32] models. The degree heterogeneity $\xi$ is tuned by the power-law exponent of species degree distribution (see SI Sec. ID, E). $S = 1200$, $\sigma = 1.0$, and $C = 0.167, 0.083, 0.016$ (i.e., $\langle k \rangle = 100, 50, 10$). (c) The typical distributions of eigenvalues for niche model networks with varying $\xi$. $S = 1200$, $\sigma = 1.0$, and $C = 0.167$. As $\xi$ increases the real part of the most positive eigenvalue grows monotonically.

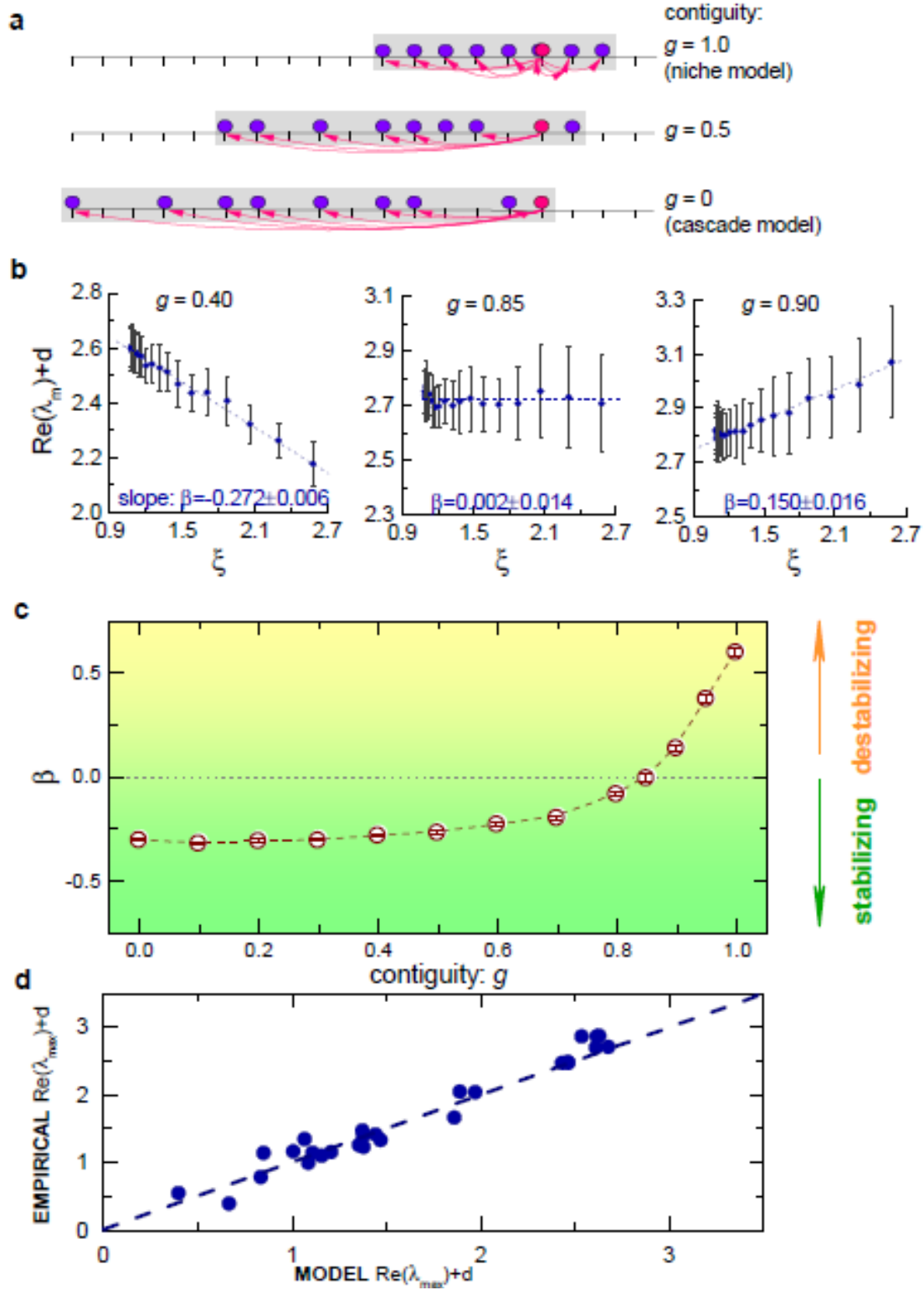

**Figure 4: Stability vs contiguity for predator-prey networks.** (**a**) Illustration of the prey contiguity $g$, which can vary from 0 to 1 and controls each predator's relative feeding range. Here, the predator (pink) consumes 8 species (purple). In the relaxed niche model (RNM) [35], when $g = 1$ the feeding range (shaded) is at its narrowest and RNM is identical to the original niche model [33] (which allows cannibalism and trophic loops). As $g$ is reduced towards zero, the feeding range (shaded) is widened while species that fall within the range have lower probability of being consumed, so that non-contiguous networks can occur (e.g., when $g = 0.5$). When $g = 0$, the feeding range is as wide as possible and RNM is equivalent to the cascade model [32, 41], i.e., the predator can consumes any species whose niche value is not larger than the predator's. (**b**) $\text{Re}(\lambda_m) + d$ as a function of degree heterogeneity $\xi$ for predator-prey networks with varying contiguity $g$. The networks are generated from the RNM. Dash lines represent the least squares linear fits. Negative slope means that degree heterogeneity is stabilizing, and positive slope indicates that increasing the degree heterogeneity will destabilize the ecological system. (**c**) Impact of degree heterogeneity on stability ($\beta$) as a function of contiguity $g$ for predator-prey networks. $S = 1200$, $\sigma = 1.0$, and $C = 0.167$. (**d**) Stability of empirical and model predator-prey networks. For each empirical network a model network is generated by the relaxed niche model whose degree heterogeneity and prey contiguity equals to that of the empirical network.

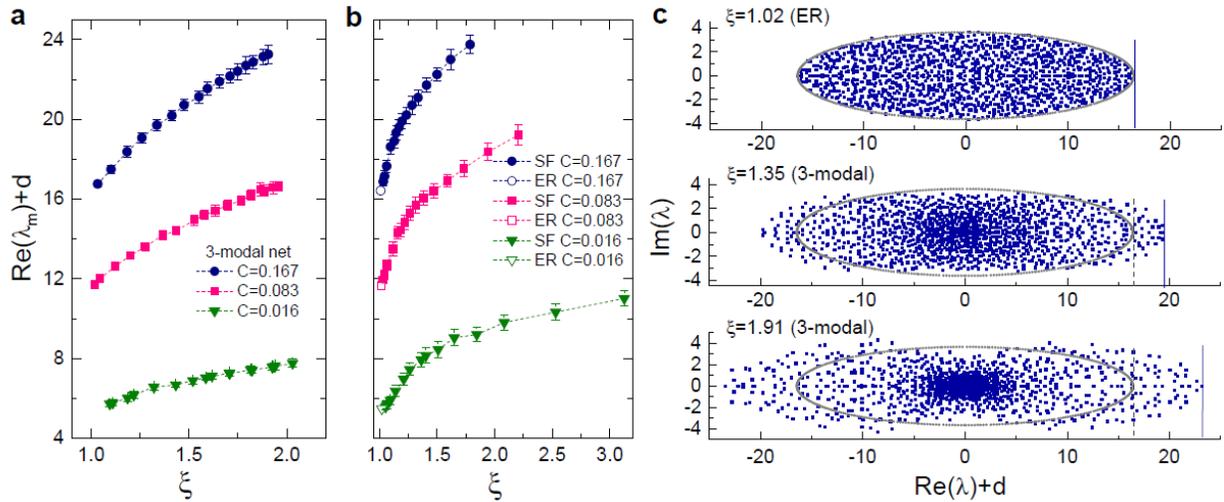

**Figure 5: Stability vs degree heterogeneity for ecological networks with mixed interactions of competition and mutualism.** (**a,b**) $\text{Re}(\lambda_m) + d$ as a function of $\xi$ for 3-modal, Erdös-Rényi (ER) and scale-free (SF) networks. $S = 1200$ and $\sigma = 1.0$. (**c**) The typical eigenvalue distribution of ER ($\xi = 1.02$) and 3-modal ($\xi = 1.35, 1.91$) networks. The ellipse (drawn for reference) is the result for homogeneous networks [14]. Despite the bow-tie shape of the eigenvalue distribution as $\xi$ increases, the real part of the most positive eigenvalue keeps increasing.